# Ising Spins on a Gravitating Sphere

*Christian Holm*[1] and *Wolfhard Janke*[1,2]

[1] Institut für Theoretische Physik, Freie Universität Berlin
Arnimallee 14, 14195 Berlin, Germany

[2] Institut für Physik, Johannes Gutenberg-Universität Mainz
Staudinger Weg 7, 55099 Mainz, Germany

**Abstract**

We investigated numerically an Ising model coupled to two-dimensional Euclidean gravity with spherical topology, using Regge calculus with the $dl/l$ path-integral measure to discretize the gravitational interaction. Previous studies of this system with toroidal topology have shown that the critical behavior of the Ising model remains in the flat-space Onsager universality class, contrary to the predictions of conformal field theory and matrix models. Implementing the spherical topology as triangulated surfaces of three-dimensional cubes, we find again strong evidence that the critical exponents of the Ising transition are consistent with the Onsager values, and that KPZ exponents are definitely excluded.

E-mail: holm@einstein.physik.fu-berlin.de
janke@miro.physik.uni-mainz.de

# 1 Introduction

The study of two-dimensional (2D) models has often proved to be an important first step for developing fundamental ideas concerning higher dimensional physics. Such has happened for 2D Euclidean quantum gravity where we have by now independent analytical results from conformal field theory and matrix models [1]. For pure gravity this is a formula for the string susceptibility exponent $\gamma_{\rm str}$, and for matter fields of central charge $c < 1$ coupled to gravity it has been found that the nature of the phase transition changes drastically. In the simplest case of the Ising spin system ($c = 1/2$), the critical exponents change from the flat-space Onsager values $\alpha = 0$, $\beta = 0.125$, $\gamma = 1.75$, and $\nu = 1$ to the KPZ [2] values $\alpha = -1$, $\beta = 0.5$, $\gamma = 2$, and $D\nu = 3$, where $D$ is the internal fractal dimension of the manifold. While these predictions could be confirmed numerically within the dynamically triangulated random surface approach, it was found [3, 4] that the Ising model on lattices of toroidal topology stays in the flat-space Onsager universality class when coupled to gravity via Regge calculus [5]. This came as a big surprise because on the basis of universality one had expected that Regge calculus, as the best understood method to discretize classical gravity, would give results compatible to KPZ scaling. The Regge method seems to depend sensitively on the path-integral measure [6], and therefore the prevailing opinion is that the measures used so far are inappropriate [7]. This issue is, however, by no means settled and still requires further investigations. It is also not clear if the commonly chosen spin coupling to gravity is the correct one.

Furthermore, in pure gravity the string susceptibility exponent $\gamma_{\rm str}$ depends on the global topology, and one might wonder if the spin phase transition also shows a dependence on the global topology. In particular the previously used torus topology plays a special role, since $\gamma_{\rm str}$ takes the classically expected value of two, suggesting the absence (or triviality) of quantum effects in this topology. In order to exclude the possibility of a fortuitous coincidence with flat-space critical exponents in the torus topology, we have therefore performed further Monte Carlo simulations of the Ising system using the lattice topology of the sphere.

## 2 The model and simulation

The model and our notation are explained in detail in Ref. [4], where we used a regular triangulation of a torus. We recapitulate here only the essential ingredients. Because in two dimension the Einstein-Hilbert action is a topological invariant, we need no curvature term. In Ref. [4] the Ising system was studied with various curvature squared terms, which showed no influence on the critical behavior of the system. Therefore in the present study we chose to have no $R^2$ term at all. This leaves only the integral over the metric $g$ to be discretized, which gives the total area $A$. We used

$$\int d^2x\sqrt{g(x)} \longrightarrow \sum_i A_i, \tag{1}$$

where the $A_i$ are taken as the barycentric areas associated with the vertices $i$. We simulated the partition function

$$Z = \sum_{\{s\}} \int \mathcal{D}\mu(l) \exp\left(-\lambda \sum_i A_i - KE(l,s)\right) \delta(\sum_i A_i - A), \tag{2}$$

where

$$E(l,s) = \frac{1}{2} \sum_{\text{edges } \langle ij \rangle} A_{ij} (\frac{s_i - s_j}{l_{ij}})^2 \tag{3}$$

is the energy of Ising spins, $s_i = \pm 1$, which are located at the vertices $i$ of the lattice, and $A_{ij}$ is the barycentric area associated with an edge $\langle ij \rangle$. The energy is the discretized analogue of the continuum action for a scalar field $\varphi$, $\int d^2x\sqrt{g(x)}g^{\mu\nu}(x)\partial_\mu\varphi\,\partial_\nu\varphi$. The delta function in (2) ensures that the total area is kept fixed at a given value $A$. As path-integral measure $\mathcal{D}\mu(l)$ we used again the simple scale invariant measure $\mathcal{D}\mu(l) = \left(\prod_{\langle ij \rangle} dl_{ij}/l_{ij}\right) F(\{l_{ij}\})$, where $F(\{l_{ij}\})$ is a function which ensures that changes in the link lengths do not violate the triangle inequalities.

As lattice topology we used the triangulated surface of a three dimensional cube of edge length $d$. This provides us with an almost regular triangulation of the sphere where six vertices have coordination number four, and all others have coordination number six. In terms of the linear length $d$ of the cube the number of vertices is $N_0 = 6(d-1)^2 + 2$. For further reference the number of links and triangles in terms of $N_0$ are given by $N_1 = 3N_0 - 6$, and $N_2 = 2N_0 - 4$, respectively. We studied ten system sizes ranging from $d = 10\,(N_0 = 488)$ up to $d = 55\,(N_0 = 17498)$. The area was kept fixed to its initial value $A = N_2/2$. The gravitational action was simulated using the standard single-hit Metropolis update. For the spin update we used the single-cluster (Wolff) algorithm [8] which prevents the critical slowing down near the phase transition. Between measurements we performed $n = 2$ Monte Carlo steps consisting of two lattice sweeps to update the link lengths $l_{ij}$ followed by one single-cluster flip to update (a fraction of) the spins $s_i$. As simulation point we chose a value of $K_0 = 1.025$, which is close to the critical coupling found in Ref. [4].

For each run we recorded the time series of the energy density $e = E/N_0$ and the magnetization density $m = \sum_i A_i s_i / N_0$. After an initial equilibration time, we performed for each lattice size about 50 000 measurements. From an analysis of the time series we obtained integrated autocorrelation times for the energy and the magnetization of about $1 - 7$ (in units of measurements) for all lattice sizes. The statistical errors were obtained by the standard Jack-knife method using 20 blocks.

From the time series we computed the Binder parameter,

$$U_L(K) = 1 - \frac{1}{3}\frac{\langle m^4 \rangle}{\langle m^2 \rangle^2}, \tag{4}$$

for each lattice of size $L = \sqrt{N_0}$, which serves as our linear scaling parameter. The curves $U_L(K)$ for different $L$ cross around $(K_c, U^*)$ with slopes $\propto L^{1/\nu}$, apart from confluent corrections explaining small systematic deviations. This allows an almost unbiased estimate of the critical coupling $K_c$, the critical correlation length exponent $\nu$, and the renormalized charge $U^*$. We further analyzed the specific heat,

$$C(K) = K^2 N_0 (\langle e^2 \rangle - \langle e \rangle^2), \tag{5}$$

the (finite lattice) susceptibility,

$$\chi(K) = N_0 (\langle m^2 \rangle - \langle |m| \rangle^2), \tag{6}$$

the susceptibility in the disordered phase,

$$\chi'(K) = N_0 (\langle m^2 \rangle), \tag{7}$$

and studied the (finite lattice) magnetization at its point of inflection, $\langle|m|\rangle|_{\rm inf}$. The inflection point can be obtained from the maximum of $d\langle|m|\rangle/dK$, which can be calculated from the time series as

$$\frac{d\langle|m|\rangle}{dK} = \langle E\rangle\langle|m|\rangle - \langle E|m|\rangle. \tag{8}$$

We further analyzed the logarithmic derivatives $d\ln\langle|m|\rangle/dK$ and $d\ln\langle m^2\rangle/dK$.

# 3 Results

By applying reweighting techniques we first determined the maxima of $C$, $\chi$, $d\langle|m|\rangle/dK$, $d\ln\langle|m|\rangle/dK$, and $d\ln\langle m^2\rangle/dK$. The location of the maxima provided us with five sequences of pseudo-transition points $K_{\rm max}(L)$ for which the scaling variable $x = (K_{\rm max}(L) - K_c)L^{1/\nu}$ should be constant. Using this information we then have several possibilities to extract the critical exponent $\nu$ from (linear) least square fits of the finite-size scaling (FSS) Ansatz $dU_L/dK \cong L^{1/\nu}f_0(x)$ or $d\ln\langle|m|^p\rangle/dK \cong L^{1/\nu}f_p(x)$ to the data at the various $K_{\rm max}(L)$. The resulting exponents $1/\nu$ from fits with an acceptable goodness-of-fit parameter $Q$ are collected in Table 1. They give mostly results which are smaller than the Onsager value $\nu = 1$, but are still compatible within the $2\sigma$ range. The relatively large deviations are probably due to the fact, that we use a non-regular triangulation, which may induce additional corrections to scaling.

Assuming $\nu = 1$ we have next determined estimates for $K_c$ from the Binder-parameter crossings and the scaling of the various $K_{\rm max}(L)$. The crossing points $K^\times$ of the curves $U_L(K)$ and $U_{bL}(K)$ with $b > 1$ approach $K_c$ as

$$K^\times = K_c + \kappa/(b^{1/\nu} - 1), \tag{9}$$

where $\kappa = \kappa(L)$ does not depend on $b$, and confluent corrections are neglected. This method, valid for large $b$, is usually the most precise one and performing all possible least-square fits to (9) with $L$ fixed we arrive at an estimate of the critical coupling $K_c = 1.023(1)$; see Fig. 1(a). Particularly interesting is that the crossings are close to their infinite volume value already on relatively small lattices. A similar observation with regards to Fisher zeros on spherical lattices was made in Ref. [9]. From the scaling of the various $K_{\rm max}$, shown in Fig. 1(b), we can obtain further estimates of $K_c$ from linear least-squares fits (assuming again $\nu = 1$). The error weighted estimate from all five sequences leads to $K_c = 1.024(2)$, consistent with the crossing value. As our final value we used in further analyses

$$K_c = 1.023 \pm 0.001. \tag{10}$$

In particular we can now extract $\nu$ also from the scaling of $dU_L/dK$ and $d\ln\langle|m|^p\rangle/dK$ at $K_c$; see Table 1. Here the errors reflect the combined uncertainties in $K_c$ and in the finite statistics.

Also the values of $U_L(K^\times)$ vary very little with lattice size. Since in view of the statistical errors it was therefore impossible to perform an infinite-volume extrapolation (based on confluent corrections), we simply computed weighted averages over various subsets of the data. The results turned out to be quite insensitive to the precise averaging prescription, and as final value we quote

$$U^* = 0.55 \pm 0.01, \tag{11}$$

where the error is a very conservative estimate. Notice that this value is significantly different from the corresponding estimate for toroidal lattices ($\approx 0.610 - 0.615$), i.e., it does depend on the global topology.

To extract the critical exponent ratio $\gamma/\nu$ we used the scaling $\chi \cong L^{\gamma/\nu} f_3(x)$ at the previously discussed points of constant $x$, as well as the scaling of $\chi'$ at $K_c$. The results are listed in Table 2. For the values of $\chi$ at $K_c$ we could not find a fit of decent quality, however a fit at $K = 1.020$ yielded $\gamma/\nu = 1.756(6)$ with $Q = 0.03$. Similar difficulties with fits of $\chi$ at $K_c$ have been observed on regular lattices as well, see e.g. Ref. [10]. All values for $\gamma/\nu$ are compatible with the Onsager value of $\gamma/\nu = 1.75$. The quality of the fit for $\chi_{\max}$ can be inspected in Fig. 2.

To extract the magnetical critical exponent ratio $\beta/\nu$ we used that $\langle|m|\rangle \cong L^{-\beta/\nu} f_4(x)$ at all constant $x$-values. Another method is to look at the scaling of $d\langle|m|\rangle/dK \cong L^{(1-\beta)/\nu} f_5(x)$. The fit results for $\beta/\nu$ and $(1-\beta)/\nu$ are collected in Table 3. Using our average value for $1/\nu$ in Table 1 we obtain the final estimate of $\beta/\nu = 0.14(2)$. This result again is compatible with the Onsager value $\beta/\nu = 0.125$.

The specific-heat exponent $\alpha$ is numerically always the hardest quantity to estimate. For the case of the Onsager value $\alpha = 0$ we expect a logarithmic divergence like

$$C(x, L) = a(x) + b(x) \ln L. \tag{12}$$

Indeed the data at the different fixed values of $x$ can all be nicely fitted with this Ansatz. In particular, for the fit of $C_{\max}$ with 10 data points shown in Fig. 3 we obtain $a = 0.12(4)$, $b = 0.361(9)$, with a total $\chi^2 = 13.9$ ($Q = 0.09$). We also did an unbiased three-parameter fit of the form

$$C(x,L) = A(x) + B(x)L^{\alpha/\nu}, \tag{13}$$

which gave us in the case of the fit of $C_{\max}$ and 10 data points $A = -5.6(15.)$, $B = 5.8(15.)$, and $\alpha/\nu = 0.05(11)$, with a total $\chi^2 = 13.8$ ($Q = 0.06$). The barely improved $\chi^2$ and the very small value of $\alpha/\nu$, consistent with zero within error bars, clearly support logarithmic scaling; compare Fig. 3 where the two fits are indeed hardly distinguishable. The large errors on $A$ and $B$ can be understood by expanding $L^{\alpha/\nu} = 1 + (\alpha/\nu)\ln L + \mathcal{O}((\alpha/\nu)^2)$, showing that in the limit $\alpha/\nu \longrightarrow 0$ the parameters of (12) and (13) are related by $A + B = a$ and $B = b/(\alpha/\nu) \longrightarrow \infty$. For comparison, we also included in Fig. 3 the best linear least-square fit with the KPZ prediction $\alpha/\nu = -2/3$ of the form

$$C(x,L) = A(x) + B(x)L^{-2/3}, \tag{14}$$

that resulted in $A = 2.09(2)$, $B = -6.9(2)$, and an unacceptable large total $\chi^2 = 57.2$ ($Q = 1.6 \times 10^{-9}$). Overall we can thus conclude that also for $\alpha/\nu$ our data is consistent with the flat-space Onsager value of zero.

# 4 Concluding remarks

We have performed a study of the Ising model on a spherical topology coupled to quantum gravity via the Regge calculus approach. Using the path integral measure $\prod dl/l$ we have found that, as on a toroidal topology, the critical exponents of the Ising transition agree with the Onsager exponents for regular static lattices, and the KPZ exponents are definitely excluded. The non-regular triangulation of the sphere seems to affect the finite-size behavior in a negative way, and one could probably obtain more accurate results, like those of Ref. [4] on the torus, by using a random triangulation of the sphere [11]. Unlike in the pure gravity case, where the global lattice topology enters in the formula for the string susceptibility exponent, it does not affect the critical exponents of the Ising phase transition.

## Acknowledgments

W.J. thanks the DFG for a Heisenberg fellowship. The numerical simulations were performed on the North German Vector Cluster (NVV) under grant bvpf01 and at the HLRZ in Jülich under grant hbu001. We thank all institutions for their generous support. Work supported in part by the EC under contract No. ERBCHRXCT930343.

# Tables

| Fit-type | $N$ | $1/\nu$ | $Q$ |
|---|---|---|---|
| $dU/dK$ at $K_{\rm max}^{C}$ | 7 | 0.90(11) | 0.11 |
| $dU/dK$ at $K_{\rm inf}^{\langle\lvert m\rvert\rangle}$ | 8 | 1.10(6) | 0.44 |
| $dU/dK$ at $K_{\rm max}^{\chi}$ | 8 | 1.14(7) | 0.70 |
| $dU/dK$ at $K_{\rm inf}^{\ln\langle\lvert m\rvert\rangle}$ | 8 | 1.21(10) | 0.85 |
| $dU/dK$ at $K_{\rm inf}^{\ln\langle m^2\rangle}$ | 8 | 1.23(10) | 0.87 |
| $dU/dK$ at $K_c$ | 8 | 0.96(10) | 0.11 |
| $d\ln\langle\lvert m\rvert\rangle/dK$ at $K_{\rm inf}^{\ln\langle\lvert m\rvert\rangle}$ | 8 | 1.13(5) | 0.53 |
| $d\ln\langle\lvert m\rvert\rangle/dK$ at $K_c$ | 8 | 0.96(7) | 0.20 |
| $d\ln\langle m^2\rangle/dK$ at $K_{\rm inf}^{\ln\langle m^2\rangle}$ | 8 | 1.13(5) | 0.31 |
| $d\ln\langle m^2\rangle/dK$ at $K_c$ | 8 | 0.96(6) | 0.25 |
| weighted average | | 1.08(5) | |

Table 1: Fitting results for $1/\nu$ using $K_c = 1.023(1)$. Column $N$ shows the number of fitted points, and $Q$ denotes the standard goodness-of-fit parameter. The average is computed by weighting each entry with its error.

| Fit-type | $N$ | $\gamma/\nu$ | $Q$ |
|---|---|---|---|
| $\chi$ at $K_{\rm max}^{C}$ | 10 | 1.74(2) | 0.07 |
| $\chi$ at $K_{\rm inf}^{\langle\lvert m\rvert\rangle}$ | 10 | 1.739(8) | 0.66 |
| $\chi$ at $K_{\rm max}^{\chi}$ | 10 | 1.745(6) | 0.42 |
| $\chi$ at $K_{\rm inf}^{\ln\langle\lvert m\rvert\rangle}$ | 10 | 1.755(9) | 0.28 |
| $\chi$ at $K_{\rm inf}^{\ln\langle m^2\rangle}$ | 8 | 1.73(2) | 0.29 |
| $\chi'$ at $K_c$ | 10 | 1.79(3) | 0.09 |
| weighted average | | 1.744(6) | |

Table 2: Fitting results for $\gamma/\nu$ using $K_c = 1.023(1)$. Column $N$ shows the number of fitted points. The average is computed by weighting each entry with its error.

| Fit-type | $N$ | $\beta/\nu$ | $Q$ |
|---|---|---|---|
| $\langle m \rangle$ at $K_{\rm max}^{C}$ | 7 | 0.07(4) | 0.17 |
| $\langle m \rangle$ at $K_{\rm inf}^{\langle \|m\| \rangle}$ | 8 | 0.15(2) | 0.41 |
| $\langle m \rangle$ at $K_{\rm max}^{\chi}$ | 8 | 0.15(2) | 0.34 |
| $\langle m \rangle$ at $K_{\rm inf}^{\ln\langle \|m\| \rangle}$ | 8 | 0.20(5) | 0.87 |
| $\langle m \rangle$ at $K_{\rm inf}^{\ln\langle m^2 \rangle}$ | 8 | 0.20(6) | 0.71 |
| $\langle m \rangle$ at $K_{\rm c}$ | 10 | 0.10(2) | 0.05 |
| average | | 0.14(2) | |
| | | $(1-\beta)/\nu$ | |
| $d\langle \|m\| \rangle/dK$ at $K_{\rm inf}^{\langle \|m\| \rangle}$ | 8 | 0.93(3) | 0.27 |
| $d\langle \|m\| \rangle/dK$ at $K_{\rm c}$ | 8 | 0.89(4) | 0.36 |
| average | | 0.92(3) | |

Table 3: Fitting results for $\beta/\nu$ and $(1-\beta)/\nu$ using $K_c = 1.023(1)$. The number of fitted points is given under column $N$. The average is computed by weighting each entry with its error.

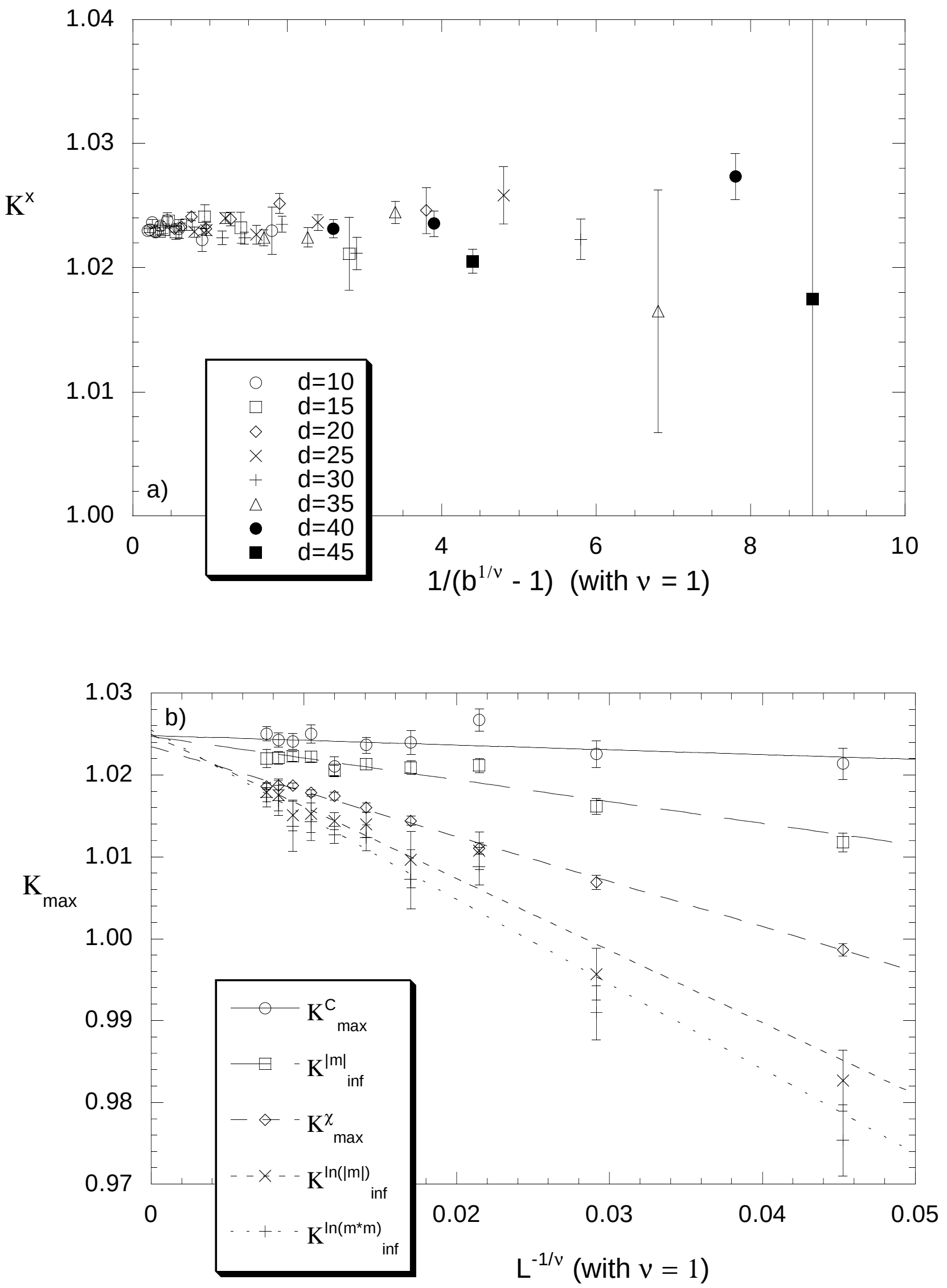

Figure 1: (a) The Binder-parameter crossings $K^{\times}(b)$ with $b = L'/L$. Assuming $\nu = 1$, straight line fits through all data points yield $K_c = 1.023(1)$. (b) Finite-size scaling extrapolations of the pseudo-transition points $K_{\max}$ vs $L^{-1/\nu}$, assuming $\nu = 1$. The error-weighted average of the extrapolations to infinite size yields $K_c = 1.024(2)$.

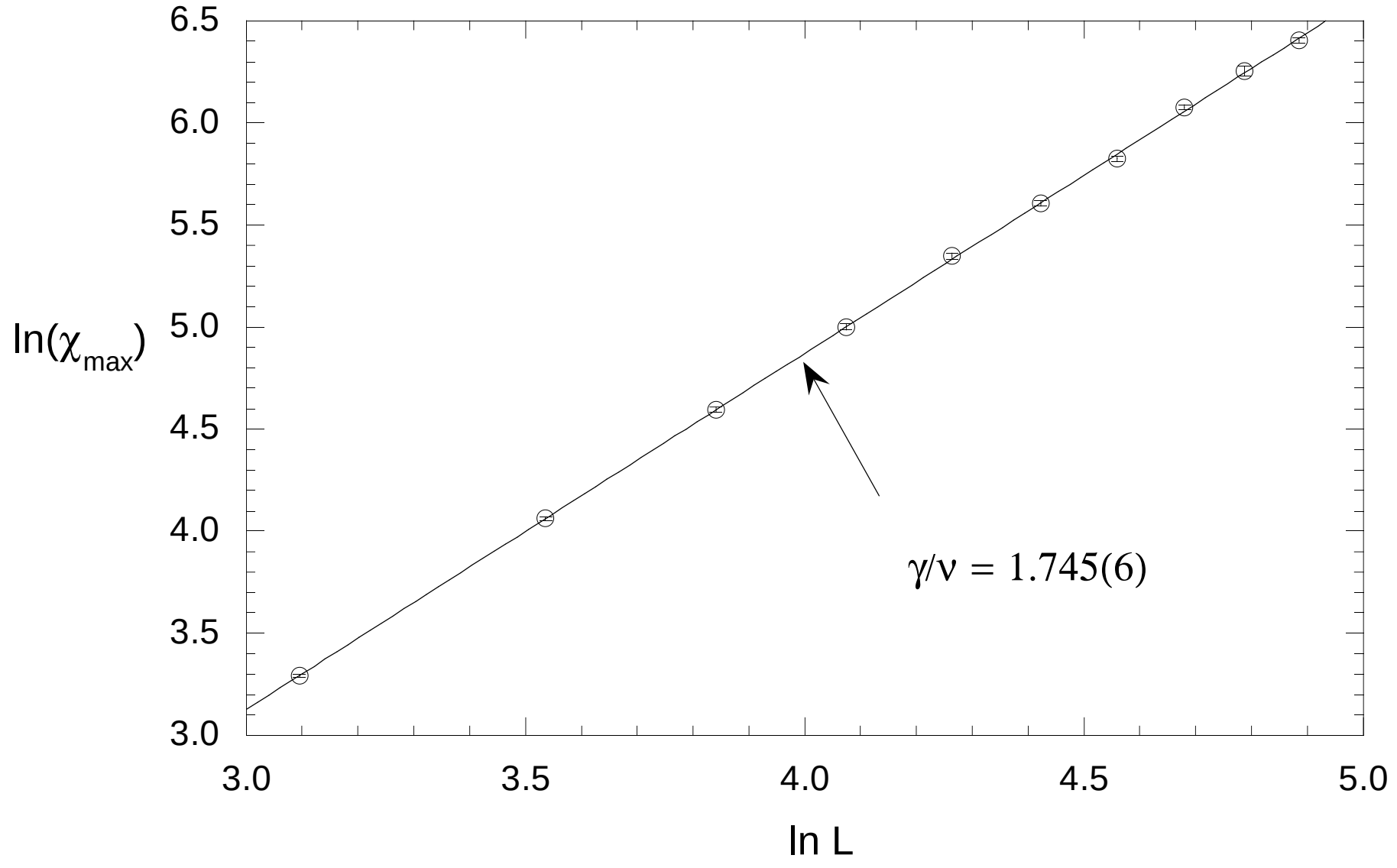

Figure 2: Finite-size scaling of the susceptibility maxima $\chi_{\mathrm{max}}$. The slope is compatible with the Onsager value $\gamma/\nu = 1.75$ for regular static lattices.

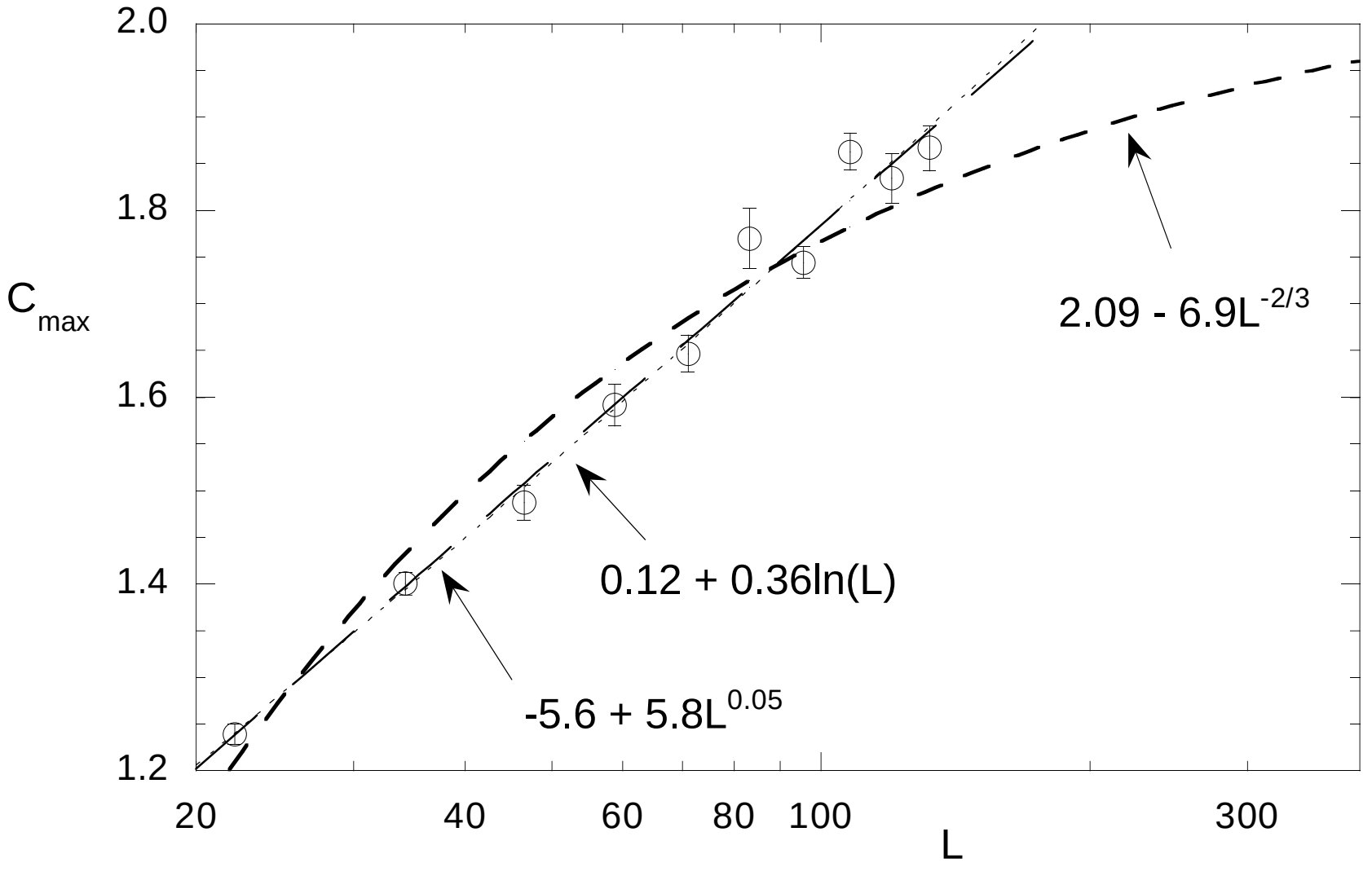

Figure 3: Finite-size scaling of the specific-heat maxima $C_{\max}$. Also shown are a logarithmic fit $C_{\max} = a + b \ln L$, a power-law fit $C_{\max} = A + BL^{\alpha/\nu}$, and a constrained power-law fit assuming the KPZ prediction $\alpha/\nu = -2/3$.